# SHIPS: A new setup for the investigation of swift heavy ion induced particle emission and surface modifications

F. Meinerzhagen[1], L. Breuer[1], H. Bukowska[1], M. Bender[2], D. Severin[2], M. Herder[1], H. Lebius[3], M. Schleberger[1] and A. Wucher[1]

[1] Fakultät für Physik, Universität Duisburg-Essen and Cenide, 47057 Duisburg, Germany
[2] GSI Helmholtzzentrum für Schwerionenforschung GmbH, 64291 Darmstadt, Germany
[3] CIMAP (CEA-CNRS-ENSICAEN-UCN), 14070 Caen Cedex 5, France

**Abstract**

The irradiation with fast ions with kinetic energies of > 10 MeV leads to the deposition of a high amount of energy along their trajectory (up to several ten keV/nm). The energy is mainly transferred to the electronic subsystem and induces different secondary processes of excitations which result in significant material modifications. A new setup to study these ion induced effects on surfaces will be described in this paper. The setup combines a variable irradiation chamber with different techniques of surface characterizations like scanning probe microscopy, time-of-flight secondary ion and neutral mass spectrometry, as well as low energy electron diffraction under ultra high vacuum conditions, and is mounted at a beamline of the universal linear accelerator (UNILAC) of the GSI facility in Darmstadt, Germany. Here, samples can be irradiated with high-energy ions with a total kinetic energy up to several GeVs under different angles of incidence. Our setup enables the preparation and in-situ analysis of different types of sample systems ranging from metals to insulators. Time-of-flight secondary ion mass spectrometry enables us to study the chemical composition of the surface, while scanning probe microscopy allows a detailed view into the local electrical and morphological conditions of the sample surface down to atomic scales. With the new setup particle emission during irradiation as well as persistent modifications of the surface after irradiation can thus be studied. We present first data obtained with the new setup, including a novel measuring protocol for time-of-flight mass spectrometry with the GSI UNILAC accelerator.

**Introduction**

The interaction of ions with solids has been of interest for many fields such as atomic physics, solid state physics and materials research. Ions can be used to modify either the surface or the bulk of the irradiated material by changing the electronic structure (e.g. implanting of ions for doping), cleaning the surface (e.g. sputter cleaning), or generation phase or structural changes e.g. by etching of irradiation damage [1], fabrication of nano pore filter [2], creation of nanoscaled surface features [3-7]. The choice of mass and acceleration voltage of the ions affect the energy density, which is deposited in a sample and offers a way to choose between different energy transfer mechanisms. At the materials research branch (M-branch) of "GSI Helmholtzzentrum für Schwerionenforschung GmbH" all stable isotopes in the periodic system of elements, from protons up to uranium can be accelerated with kinetic energies between 3.6 MeV/u up to 11.4 MeV/u with the universal linear accelerator (UNILAC). For these swift heavy ions (SHI) the nuclear stopping imparted through elastic collisions onto the target atoms plays a minor role and will only come into play, after the projectile has been slowed down. Thus, this interaction affects the bulk material at a depths in excess of several µm but not at the surface itself. The main process of slowing down SHI in matter occurs by electronic excitation and ionization processes, and is termed electronic stopping. One channel of energy dissipation from the resulting electronic excitation is given by electron-phonon-coupling. Depending on the strength of this coupling and the ion induced initial electron temperature, the solid can undergo phase transitions such as melting in a small cylinder located along the trajectory of the ion with a diameter of the order of nanometers. At the surface, this so-called thermal spike [8-10] may result in the formation of small hillock-like structures [6], a phenomenon which has been observed in many materials. Nevertheless, only little is known about the exact structure and stoichiometry of these ion induced features, and the direct observation of their formation is not possible. The time scales of this process are in the range of femto- to picoseconds and are therefore too fast to be followed by established experimental methods. But not only the details of hillock formation remained elusive. During irradiation material from the surface may be emitted into the vacuum. While the energy and angle distribution of this sputtered material has been studied for years and is relatively well known [11-13], little is known about the ionization probability of the sputtered particles. Hitherto no attempt has been made to directly correlate particle emission with hillock formation. Early experiments regarding sputtering with SHI assumed a direct connection between track formation and particle emission from solid surfaces [14-16]. For insulating materials the formation of tracks as well as the particle emission can be explained by a so called Coulomb explosion [16], where the highly ionized bulk material along the trajectory of the projectile leads to a strong electric repulsion causing material damage and sputtering. Other models as the thermal spike model use the temperature of the electronic and nuclear subsystem as the crucial parameter to model track formation and emission of material into the gas phase. In this model, the SHI stores its energy in the electronic subsystem, which provides the energy to heat the lattice subsystem. Thereby the temperature of the lattice has to exceed a critical temperature (like the melting or sublimation temperature), otherwise no track formation can be observed. The formation of surface tracks shows this threshold behavior [6,17,18] and therefore it is proved to assume that electronic excitations play an important role in their formation. However, surface tracks are usually investigated after irradiation by scanning probe microscopy where the information is basically limited to permanent morphological modifications. These measurements provide e.g. important threshold values but they do not provide neither any details about the inner structure of the tracks as e.g. their stoichiometry, nor do they give information about transient modifications. Surprisingly, almost all surface track phenomena could successfully be described in terms of the thermal spike, see i.e. [3,6,19] . However, even if this was true this would not exclude the possibility of other processes taking place and contributing to the final state. By analysing only the final modification their possible contribution may not be resolved. Therefore, direct insight into the exact nature of the energy dissipation mechanisms is difficult to obtain and thus complementary methods are required. As it is demonstrated here, this can be achieved by mass spectrometry. In particular, by analysing the ejected particles from a compound material in-situ during the irradiation one may be able to distinguish between different models: Coulomb explosion e.g. would yield a distribution of ejected charged particles corresponding to their respective ionization probabilities, whereas a thermal process would give rise to a distribution corresponding to the stoichiometry of the compound material. Especially for compound materials like $SrTiO_3$ it is therefore interesting to investigate the composition of the sputtered material during irradiation, as the detailed analysis of the emitted material provide insight into the ejection mechanism and may give evidence to chemical structure of the hillock-like structures. The latter may be revealed by connecting the experimentally determined differential sputtering yield and the surface modifications with a special emphasis on hillock formation. Such an analysis may be performed with the setup presented here, where two complementary surface science techniques, namely high-resolution SPM under ultra high vacuum (UHV) conditions and time-of-flight secondary ion and secondary neutral mass spectrometry (TOF-SIMS/SNMS), are combined with the UNILAC beamline for the first time. With our setup it is possible to obtain complementary information on ion induced surface modifications as a function of irradiation parameters such as energy and angle of incidence. In the following we will first describe the respective methods and their implementation in detail and will then proceed to present first data that demonstrates the potential of our new setup, called SHIPS (the acronym for Swift Heavy Ion induced Particle Emission and Surface Modifications).

**Experimental setup**

The experimental setup presented here has to fulfill different requirements for the intended experiments. The irradiation of samples with SHI implicates the use of an accelerator that provides a total energy of more than 10 MeV - in our case the M1-branch at the UNILAC beamline at the accelerator facility in Darmstadt. An important parameter of the experiment is the positioning of samples. Height and angle of the sample surface with respect to the ion beam have to be adjusted very carefully within the range of some millimeters to position the SHI-beam, a keV-beam, a laser, and the Time-of-Flight (TOF) spectrometer along the sample and, depending on the intended experiment, with a precision of a tenth of degree for irradiations under grazing angle of incidence.

Detailed studies of nanosized surface modifications necessitate a microscope with sufficient resolution, like an atomic force microscope (AFM), a scanning tunneling microscope (STM) or a scanning electron microscope (SEM). Here, we use a scanning probe microscope (SPM) combining AFM and STM. These methods provide the required resolution, STM and Kelvin probe give access to the local electronic properties of the surface, and AFM allows to probe also electrically insulating samples.

For the analysis of the sputtered particles, mass spectrometric methods are used. To investigate small mass particles as well as huge organic molecules with masses up to several hundred atomic mass units, we use a TOF spectrometer, which allows the parallel detection of ions of different masses, thereby greatly enhancing the detection sensitivity with respect to a mass filter. Normally, only the ionized part of the emitted material ("secondary ions") is directly accessible to mass spectrometric detection and the resulting technique ("Secondary Ion Mass Spectrometry SIMS") is well established as a standard method for chemical surface analysis. Unfortunately, the ion fraction of the sputtered material is in many cases small and may in addition critically depend on the chemical environment of the ejected particles. For quantitative analysis, and in order to gain insight into the sputtering mechanism itself, those particles which are emitted in the neutral charge state ("secondary neutrals") therefore need to be investigated as well. In the setup described here, we therefore combine SIMS with its neutral counterpart ("Secondary Neutral Mass Spectrometry SNMS"), which is performed by post-ionization of secondary neutral particles via single photon absorption in an intense, pulsed vacuum-ultraviolet (VUV) laser directed closely above and parallel to the irradiated surface. After post-ionization, the TOF spectrometer cannot distinguish between secondary ions and the corresponding post-ionized neutrals, thereby allowing a direct comparison of SIMS and SNMS signals in order to determine the ionization probability of a sputtered atom or molecule. While this strategy has been previously employed to investigate secondary ion formation in the nuclear sputtering regime (i.e., under bombardment of elemental surfaces with atomic ions of keV impact energies), it is here for the first time implemented to investigate the ion fraction of material ejected under electronic sputtering conditions. Detailed knowledge of this quantity is crucial to improve our understanding of the sputtering mechanism under such conditions, but mass resolved data of this kind is practically nonexistent. In order to align the instrument and facilitate a comparison with nuclear sputtering, the setup is equipped with a low energy ion source delivering a beam of rare gas ions with energies up to 5 keV. Both ion beams impinge onto the surface under the same impact angle (45° with respect to the surface normal), while the TOF spectrometer detects sputtered particles emitted along the surface normal. This way, the same experimental conditions are used under MeV/u and keV sputtering, thereby enabling the quantitative measurement of ionization probabilities under MeV/u bombardment via a direct comparison with similar data previously measured under keV bombardment.

The preparation of clean surfaces and their investigation with a time-of-flight mass spectrometer requires an ultra-high vacuum (UHV) environment. Also, the transfer process for samples has to be fast enough, so that contamination of the surface by adsorbed residual gas is kept to a minimum. This is of particular importance for reactive surfaces, e.g. clean, reconstructed silicon surfaces or metal surfaces without an oxide layer.

*(I) Vacuum and transfer system*

The UHV system (for an overview see figure 1) was built at the terminal of the M1 beamline. It thus presents a major extension of the beamline described in Ref.[20]. The setup includes four main chambers: the load lock (a) for the transfer to vacuum, the preparation chamber (b) for systematic manipulation of samples, AFM/STM chamber (c) with the SPM for characterization of samples down to atomic scales, and the irradiation chamber (d), where samples are irradiated with the ions delivered by the UNILAC and are simultaneously characterized in-situ by TOF-SIMS/SNMS. The beam diagnostic chamber (e) is part of the ion beamline and important for ion beam characterization. All chambers are mechanically decoupled by bellows to avoid disturbing mechanical vibrations during SPM measurements.

*(a) Load lock*

Samples as well as Probes for the SPM can be transferred to the UHV via the load lock chamber, which is separated from the rest of the vacuum system by a manual gate valve. A transfer rod with a specially designed transfer adapter (Ferrovac GmbH, see figure 2) allows the introduction of five carriers (either for probes or for samples) at once. All positions are designed for the Omicron sample holder design. The first sample position of this transfer adapter is tilted by an angle of 45°. This sample position is required for the transfer to the manipulator in the preparation chamber (b). The pressure is monitored by a wide range vacuum gauge (Extraktor IE514, Leybold).

*(b) Preparation chamber*

The preparation chamber includes different options for the preparation of samples and their subsequent analysis. The central part of the chamber is the manipulator (VG Scienta). The sample can be transferred with a wobble stick (Ferrovac GmbH) from the first transfer adapter position to the manipulator (see figure 2). Two different heating options are available at the manipulator: a direct current heating (e.g. for flashing of silicon samples) and a resistive heater, a boron nitrite plate with a live wire for indirect heating. The temperature is detected by thermo couples close to the sample holder of the manipulator and by a pyrometer pointed at the sample surface through a window of the chamber. A dual electron beam evaporator (tectra GmbH) in the chamber allows the deposition of two different materials onto the surface, with the mass flow being controlled by a quartz crystal microbalance. The surface can be dosed with various gases (e.g. hydrogen, oxygen etc) via a manually controlled high precision gas inlet valve. A micro-channel plate low energy electron diffraction unit (MCP-LEED, Schaefer Technologie GmbH) in the chamber can be used to analyze and control the surface crystallinity. The use of MCPs allows for extremely low beam currents in the order of pA, which means that the LEED system can also be used in the case of poorly conducting surfaces.

*(c) AFM/STM chamber*

The Variable Temperature UHV SPM (Omicron NanoTechnology GmbH, Taunusstein) is located in the SPM chamber (see figure 1). For the transfer of samples and SPM probes a wobble stick is installed. Up to twelve probes or samples can be stored in a carousel within this chamber. The wobble stick is used to move the carrier from the transfer rod coming from the load lock to the carousel or the measuring unit of the SPM and allows also moving of samples to a second transfer rod for the transfer to the irradiation chamber. Other components in the chamber are a turbo pump (Pfeiffer), an ion getter pump with titanium sublimation pump (300L, Gamma-Vacuum), and a vacuum gauge to measure the pressure in the chamber. The base pressure here is $5 \cdot 10^{-10}$ mbar.

*(d) Irradiation chamber*

The irradiation chamber forms the end of the UNILAC M1 beamline (see figure 1). It houses the sample stage for the irradiation, the home-built TOF-MS (construction drawings see [23]), a keV-sputter ion gun (IQ-100, Leybold), an *xyz*-manipulator holding a $CaF_2$-lens for focusing a VUV-laser beam onto the detection volume, a photoelectric detector to monitor the laser pulse and intensity, a residual gas analyzer (RGA) and a vacuum gauge (see figure 3a for the allocation of the respective flanges). A second transfer rod from the AFM/STM to the irradiation chamber is used to transfer one of the sample holders into the irradiation chamber, where it is then positioned on the irradiation stage via a third wobble stick.

The irradiation stage allows positioning the sample at variable height and angle with respect to the MeV ion beam. This is achieved by two linear motion drives with a lifting range of around 50 mm, which are operated by software-controlled stepper motors. The height of the stage is monitored and adjusted by a height and line-of-sight calibrated telescope, which is also used for the alignment of the complete M1 beamline. For irradiation under grazing angle of incidence, the tilting angle of the sample on the stage is calibrated and adjusted by the reflection of an alignment diode laser located on the top window of the chamber (see figure 3b). Depending on the size of the laser spot on the sample and the surface reflectivity, a resolution of 0.5° is possible in the range of impact angles between 5° to -5° with respect to the surface. The geometry of the top flange restricts the light path of the diode laser. For impact angles >5°, the laser does not exit though the top window again, so these angles have currently to be adjusted with much less precision by using the stepper motors without additional calibration. The range of angles provided from the stage geometry and stepper motors is from around 60° to -2°. For TOF-SIMS/SNMS experiments, the sample stage is tilted to an angle of 45° relative to the UNILAC beam (see figure 3c), with the surface normal directed along the ion optical axis of the mass spectrometer. Here, the adjustment of the correct angle is done by observing the measured signal. An additional specification for the stage is its electrical insulation against the chamber to enable the ion extraction into the TOF spectrometer by pulsing the sample from ground potential to high voltage.

During the build-up phase of the experiment, the keV ion beam was used to align the instrument and find correct settings (time delays and voltages) for the TOF-MS without having to run the UNILAC. For that purpose, the rare gas ion source was mounted such that its beam impinges under roughly the same impact angle with respect to the surface normal as the UNILAC beam. If both beams are directed to the same surface area, the instrument therefore cannot distinguish between particles sputtered from the surface under impact of the UNILAC and the keV ion beam, respectively. In the standard operation protocol of a TOF-SIMS/SNMS experiment, the primary ion beam is operated in a pulsed mode with pulse durations of a few nanoseconds up to a few microseconds. Running with 5 keV acceleration voltage and argon gas, the gun allows a pulsed operation of the beam with pulse lengths ranging from 100 ns to infinity, a feature which will become important for the success of the experiment (see details in section II).

For post-ionization of sputtered neutral particles, an $F_2$ excimer laser (ATLEX-500-l, ATL Lasertechnik GmbH) operated at a wavelength of 157 nm and a maximum repetition rate of 500 Hz is used. The laser beam is guided to the vacuum chamber via an evacuated beamline held at a pressure of about $10^{-3}$ mbar, which contains an adjustable 62.5° deflection mirror and a $CaF_2$ focusing lens of 300 mm focal length. The lens is mounted on a *xyz*-manipulator for beam alignment and also forms the vacuum window to the UHV system, so that the evacuated beamline acts as a differential pumping stage to minimize the leak rate across the O-ring seal between the lens and the UHV chamber. The laser provides VUV pulses with 5-8 ns duration and a maximum energy per pulse of 1.8 mJ, which is monitored by an internal energy detector mounted in the laser itself. The translation of the lens via the *xyz*-manipulator is used to adjust the position as well as the focal size of the laser beam within the ionization volume located about 1 mm above the sample surface.

Within the irradiation chamber, the laser pulses are monitored by a photoelectric detector sketched in figure S1 of the supplementary information. The detector is located on the bottom side of the irradiation chamber opposite to the laser entry flange (see figure 3). It consists of a 90% transmission wire mesh grid and a gold coated stainless steel collector plate, which are separated by a gap of about 1 mm. While the grid is kept at ground potential, a voltage of about 1.2 kV is applied between grid and collector in order to accelerate photoelectrons emitted from the grid to the collector, thereby generating a current pulse which is monitored via the voltage drop across a 50 Ω resistor placed in the grid line (see figure S 1).

This way, only a relatively small part of the laser photon flux is utilized to generate the photoelectron signal, thereby minimizing the space charge between grid and collector. The detector signal is used to monitor the timing of the laser pulse as well as the pulse energy of the laser beam after passing the optical components on a shot-to-shot basis.

For the control of the vacuum conditions in the chamber, there are a residual gas analyzer (MKS Microvision 2, MKS Instruments Deutschland GmbH) and a vacuum gauge. A second gauge is installed close to the gas dosing valve in the keV sputter gun.

*(e) Beam diagnostic chamber*

The beam diagnostic chamber includes a Faraday cup that is moved by a remote controlled pneumatic linear drive for the detection of the current from the UNILAC ion beam. The Faraday cup in the beam diagnostic chamber is the second and last one in the M1 line of the M branch of GSI. In combination with an aperture inside the beamline, the fluence during the irradiation can be monitored. A detailed description of this part of the UNILAC can be found in [20].

*(II) Scanning probe microscope*

The SPM is equipped with a beam deflection AFM and STM options. The piezo scanners are designed for a scan range *x-y-z* of 10 µm x 10 µm x 1.5 µm with a *z*-resolution of < 0.01 nm and a lateral resolution less than 1 nm, providing atomic resolution. A long focal length optical microscope is used for navigating the probe across the surface, for the approach procedure of the SPM probe, and for probe exchange. The beam deflection AFM is usually operated with silicon or $Si_3N_4$ cantilevers with nanometer sized pyramidal tips as a force sensor. A light spot from a laser diode is placed on the backside of the cantilever and reflected onto a segmented photodiode, which is then used for the detection of the tip-surface-interaction induced deflection of the cantilever. Two different operation modes, the static contact mode and the dynamic non-contact mode, are possible. Note that in the latter mode, in contrast to conventional ambient atomic force microscopes, the oscillating cantilever is always operated at its resonance frequency. Therefore the feedback signal in this mode is the frequency shift *df* and not the amplitude as in conventional tapping mode. This frequency modulated (FM) mode is directly connected to the extremely high Q-factor of the oscillating probe and warrants the highest possible resolution as well as an unambiguous distinction between conservative and non-conservative interactions [21]. By choosing different kind of tips and employing various feed-back and detection schemes, the microscope can be operated to detect e.g. electrostatic forces, contact potential differences or magnetic interactions.

The STM with metal tips as sensors can be operated at low currents (< 1 pA) but can be used also in applications with high currents (up to 330 nA). The preamplifier is equipped with a software controlled gain switch, a maximum bandwidth of 80 kHz and an optional compensation for voltage offset and tunneling current offset. The possible gap voltages range from ± 5 mV up to ± 10 V and are applied to the tip and cantilever, respectively, while the sample is grounded.

### *(III) Time-of-flight spectrometer*

The setup of the SIMS/SNMS-experiment consists of a reflectron type time-of-flight spectrometer mounted at an angle of 45° with respect to the UNILAC beamline (see figure 3). During the TOF experiments, the relative orientation of the sample is arranged such that the ion optical axis of the TOF spectrometer is perpendicular to the sample surface, so that secondary ions as well as post-ionized sputtered neutrals are extracted and detected along the surface normal. The rare gas ion gun is mounted on a flange that points to the center of the irradiation chamber under an angle of 45° with respect to both the UNILAC beam and the vertical axis. The keV ion beam therefore impinges onto the sample surface under an impact angle of 31.4° with respect to the surface normal, with its plane of incidence being rotated by 45° with respect to that of the UNILAC beam. The post-ionization laser beam enters the chamber via a second 45° flange as indicated in figure 3, so that its beam direction forms an angle of 81.6° with the sample surface normal. As a consequence, the laser beam traverses the sample under an angle of 8.4°, i.e., nearly parallel to the sample surface.

### *(a) Design of the spectrometer*

The TOF-MS spectrometer is based on a homemade design described in detail in [23]. A schematic drawing of the spectrometer and its applied voltages is shown in figure 4.

The spectrometer consists of an extraction optics comprising two entrance electrodes ($U_{extraction\ I\ and\ II}$ in figure 4), which are normally both kept at ground potential, followed by an electrostatic lens ($U_{lens}$ in figure 4) to focus the extracted ion beam onto the detector. Following the lens, deflecting plates ($U_{left/right}$ and $U_{forward/backward}$ in figure 4) are used to steer the ions through the spectrometer. It is noteworthy that the deflector settings applied here can be used to move the sensitive volume of the spectrometer, i.e., the volume from which ions are extracted and detected and therefore contribute to the measured flight time spectrum, in directions parallel to the sample surface. When the ions have passed the extraction optics, they traverse a field-free drift region, before entering a two stage reflectron consisting of a retarding ($U_{retard}$ in figure 4) and a reflection field ($U_{reflect}$ in figure 4). After reflection, the particles again enter the field free drift zone before reaching the detector. As described in detail elsewhere, the settings of $U_{retard}$ and $U_{reflect}$ determine the flight time focusing conditions of the spectrometer. By setting $U_{reflect}$ lower than the stage potential during ion extraction, ions originating from a position located below a minimum height above the surface can be prevented from being reflected and, hence, reaching the detector. By a proper selection of the $U_{retard}/U_{reflect}$ ratio, flight time focusing to first order is established for a chosen extraction height above the surface, which then determines the vertical location of the sensitive volume.

The detector consists of two micro channel plates (MCPs) in chevron configuration. In difference to [23], the setup has been redesigned to allow post-acceleration by up to 10 kV ($U_{acceleration}$ in figure 4) in order to boost the impact energy of the analyzed ions onto the MCP and enhance the detection efficiency particularly of larger molecules. The gain voltage across the MCP ($U_{gain}$ in figure 4) is varied between 1.7 and 2.25 kV depending on the measured signal level. The electrons produced at the backside of the MCP are extracted onto the collector by a voltage of +45 V, and the resulting signal is AC coupled to the data acquisition hardware installed in the computer controlling the instrument via a high voltage capacitor followed by a 2x preamplifier.

Data acquisition is possible in two different modes; First, a transient digitizer (TD) (PX1500-2-(AMP/XF)-M, Signatec) can be used to record the entire trace of the measured signal with a time resolution of 1 ns and a maximum repetition rate of 1 to 2 kHz, depending on the length of the recorded TOF spectrum. Second, a multi-stop time to digital converter (TDC) (P7888, Fast ComTec) can be used for single ion counting at a time resolution of 1 ns with practically negligible dead time between successive stops and sweeping rates of up to 10 kHz. While the TDC mode is restricted to spectra containing only one ion of each mass in each sweep, the TD mode is suitable to record detector signals generated by more than one ion impinging at the same time and may therefore record a complete TOF spectrum in a single sweep. As shown below, this feature is particularly important in SNMS operation, where the signal recorded in a single spectrum may arise from many ions of the same mass that are detected simultaneously.

*(b) Timing in SIMS and SNMS*

In a TOF-SIMS experiment, the sputtered ions are accelerated by an extraction potential ($U_{target}$ in figure 4) of about 1.5 kV. As a result, they enter the field-free drift zone with nearly the same kinetic energies and are mass separated by measuring their velocity via the flight time to their arrival on the detector. Since this measurement needs a precise starting time, either the ion generation or their extraction into the spectrometer needs to be performed in a pulsed mode, thus making the timing of the experiment a crucial parameter.

In principle, two approaches can be followed. In "DC extraction" mode, the sample is constantly kept on high voltage potential, leaving the ion extraction field permanently on. In that case, the *generation* of secondary ions defines the time resolution by using a pulsed primary ion beam with relatively short ion pulse lengths of the order of nanoseconds. In the "delayed extraction" mode, on the other hand, the sample is kept at ground potential during the ion bombardment, and the time resolution is defined by their *extraction* via fast switching the sample potential to high voltage. Under these conditions, the duration of the primary ion pulse is irrelevant for the flight time (or mass) resolution and can therefore be increased in order to boost the measured signal. Typically, ion pulse duration of a few microseconds is sufficient to fill the sensitive extraction volume with sputtered particles of all emission velocities, leading to a saturation of the measured signal with no effect of a further increase of the primary ion pulse width. The extraction potential is usually fired shortly (a few ns) after the ion pulse in order to maximize the measured signal and ensure that the primary ion beam does not bombard the surface while the extraction field is switched on.

For the analysis of sputtered neutrals, an additional complication arises from the pulsed nature of the post-ionization laser. Usually, the laser pulse is fired shortly before the extraction potential, so that the post-ionized neutrals are extracted in the same way as secondary ions present in the sensitive volume. As long as the extraction field is switched off during ion bombardment, secondary ions emitted from the surface expand freely into the vacuum in the same way as their sputtered neutral counterparts. If the laser beam is properly aligned to intersect the sputtered plume exactly in the sensitive volume of the TOF spectrometer, the instrument cannot distinguish between secondary ions and post-ionized neutrals of the same species, thereby rendering the experimental conditions such as their collection efficiency, transmission and detection efficiency exactly identical. Provided the post-ionization efficiency is known, quantitative information about the ionization probability, or ion fraction, of a sputtered species (atoms or molecules) can therefore be obtained from a direct comparison of SIMS-spectra (taken without the laser beam) and SNMS-spectra (taken with the laser pulse fired).

Due to special conditions stipulated by the accelerator facility, new concepts for the timing of the experiment had to be developed, which are shown schematically in figure 5. The ion pulse provided by the UNILAC has a width of 1 to 3 ms depending on the ion source type used by the GSI. The maximum repetition rate of the accelerator is 50 Hz, but, the generated ion pulses are usually divided between different end user experiments by kicker magnets, rendering the typical repetition rate usable for our experiment of the order of several Hz. In the context of a TOF spectrum acquisition, these conditions correspond to a delayed extraction mode with quasi-dc primary ion bombardment, where the sample is continuously bombarded even during the time when the extraction field is switched on. Under the prevailing operation conditions, this does not harm the experiment, since secondary ions being desorbed from the surface during that time experience the full extraction potential and are therefore not reflected and detected in the TOF spectrum. Using the standard TOF-MS protocol, the acquisition of mass spectra would under such conditions encompass a series of relatively slow data acquisition cycles ("sweeps"), where one extraction pulse would be fired at some time during or shortly after each UNILAC pulse, leading to an experiment which is in standby during most of the time. In addition, each mass spectrum acquired with the UNILAC beam needs to be complemented by a blank spectrum taken with the UNILAC off in order to unambiguously identify the signal induced by the SHI bombardment.

In order to save valuable beam time, an interleaved data acquisition protocol has been developed to improve the usage of the pulses of the UNILAC. As a first step, the extraction pulse is fired several times within each UNILAC ion pulse in order to collect multiple sweeps within the same pulse (see figure 5c). For that purpose, the delay generator controlling the TOF data acquisition timing (Model 588-1U-8C, Berkeley Nucleonics Corporation) is programmed to run at the maximum possible repetition rate (1.2 kHz depending on the length of the recorded flight time interval) and gated with a trigger signal that is generated by the accelerator facility and mimics the UNILAC pulse (in the following referred to as "macropulse"). This way, several sweeps can be acquired within one single primary ion pulse, thereby greatly reducing the number of UNILAC pulse cycles required to accumulate a certain number of sweeps in order to reach a TOF spectrum with acceptable statistics.

As a second step, it appears straightforward to use the relatively long pause between subsequent UNILAC pulses in order to acquire the corresponding blank spectra, i.e., spectra taken without ion bombardment under otherwise identical experimental conditions. For that purpose, an electronic circuitry was implemented (in the following referred to as "switchbox"), which measures the length of the UNILAC pulse (blue in figure 5) and provides an additional gating pulse of the same length for the delay generator (green in figure 5). This second gate pulse is programmed to follow each UNILAC pulse at a delay of five times its width, a feature which is important since the UNILAC pulse length may in principle change on a pulse to pulse basis without notice. This way, the blank spectra are collected within the same UNILAC pulse cycle, thereby cutting the required beam time to half.

As outlined above, one of the fundamental goals of this experiment is to compare the spectra collected with the MeV/u UNILAC ion beam with those collected with 5 keV $Ar^+$ primary ions, again under otherwise identical experimental conditions. For that purpose, another gate pulse is generated (orange in figure 5), which enables data acquisition and at the same time switches the argon gun blanking voltage (see $U_{blanking}$ in figure 4) off. Again, this third gate pulse is fired at a delay of 5 times the UNILAC pulse width after the end of the blank gate window. The width can be adjusted for optimum usage of the break between two subsequent UNILAC pulses. Since the keV ion beam is continuously switched on during this gate, a series of TOF spectra is accumulated under quasi-dc keV sputtering conditions, which are directly comparable to those acquired under MeV/u bombardment. Alternatively, it is also possible to work with a pulsed keV beam with pulse lengths of typically 2 µs, as indicated in figure 5.

In SNMS mode, an additional limiting factor is the post-ionization laser, which permits a maximum pulse repetition rate of 500 Hz. In order to still be able to use the maximum data acquisition rate in every gate window, the laser is therefore triggered only for the first sweep of each gate, leaving the remaining sweeps for the corresponding SIMS acquisition. For that purpose, the switchbox only routes the first laser trigger generated by the delay generator actually through to the laser, as shown in figure 5d. This way, a total of six different TOF spectra can be acquired quasi-simultaneously during one single UNILAC pulse cycle.

In order to get this interleaved data acquisition protocol to work, the software controlling the experiment needs to properly sort the different sweeps into the appropriate TOF spectra. When the laser is fired and at the same time the sample is bombarded with the UNILAC pulse, the measured signal contributes to the MeV-SNMS spectrum. For all other spectra within the same MeV gate window, the laser trigger pulse is suppressed and the data are added to the corresponding MeV-SIMS spectrum. Within the following gate window, the surface is neither bombarded with keV ions nor MeV/u ions, and the first sweep collected with the laser being fired contributes to a background spectrum produced by the laser alone via photoionization of residual gas atoms or molecules. The remaining sweeps within this blank gate window then contribute to the control blank spectrum, which should basically reveal the noise limited baseline. In the following keV gate window, the ion gun trigger produced by the delay generator is routed to the switching unit controlling the blanking of the Argon beam, leading to pulsed $Ar^+$ ion bombardment of the sample as indicated in figure 5d. Alternatively, the keV ion beam can be switched on permanently during this gate window as described above. The first sweep in the gate window is again an SNMS cycle, and the data is therefore added to the keV-SNMS spectrum, while all remaining sweeps contribute to the keV-SIMS spectrum, respectively.

Synchronization of this hardware generated protocol with the software running the experiment is tricky, since the computer has no control over the trigger of either the UNILAC or the delay generator. Under these conditions, the software arms the data acquisition hardware for a sweep and then waits for an "acquisition complete" flag, indicating that the selected data acquisition board (TD or TDC) has received a start trigger and completed a sweep before reading the data. Since the switchbox routing the pulses generated by the delay generator to the respective hardware components is triggered by the UNILAC macropulse, the first sweep recorded after starting a measurement necessarily corresponds to MeV-SNMS, while those following within a time window of about 5 ms (the maximum possible UNILAC pulse length) correspond to MeV-SIMS, respectively. The first sweep recorded after the end of that time window then again corresponds to SNMS (this time the residual gas spectrum); while all sweeps following this one in another time window of about 5 ms correspond to the blank spectrum. The next sweep recorded after the end of the blank time window then corresponds to keV-SNMS, while the ones following within the selected keV gate width (which needs to be known to the software) correspond to keV-SIMS.

In our first attempt to realize the interleaved data acquisition protocol, the delay generator was operated in (gated) free running mode without any synchronization to the UNILAC pulse. In that case, the entire pulse train depicted in figure 5c-e travels across the gate windows from pulse to pulse, unless the repetition rate used during data acquisition exactly matches an integer multiple of the UNILAC pulse repetition rate. Since the latter is not under our control and, moreover, may be subject to sudden changes during an experiment, this protocol did not result in satisfactory data. As a consequence, the delay generator was switched to operate in burst mode and triggered by the gate output pulses generated by the switchbox, with each trigger generating a preselected number of data acquisition cycles (sweeps) with a preselected repetition rate. Again, the unit is gated such that only sweeps within the three different gate windows are allowed while the remaining cycles of a burst are blocked. This way, the data acquisition sweep chain is fixed within each gate window, resulting in stable timing conditions with a constant number of sweeps occurring in each gate. In order to precisely control the relative timing of the UNILAC and data acquisition, another delay was introduced between the UNILAC macropulse and the actual pulse triggering the bursts, so that the entire data acquisition pulse train can be shifted within the gate windows. The usefulness of this feature will be described below.

In some cases, it has proven useful to suppress the bombardment of the investigated surface with the keV $Ar^+$ ion beam in order to avoid ion induced damage of the sample. This is particularly important for the analysis of molecular samples, where it is well known that bombardment with atomic projectiles at keV energies often results in significant fragmentation, which may accumulate to such an extent as to completely destroy the molecular information contained in the mass spectrum. On the other hand, emission of intact molecules has been observed under SHI bombardment, suggesting that MeV-SNMS/SIMS analysis of such samples may be possible without significant damage accumulation. Therefore there is an option to suppress all trigger pulses for the keV ion gun.

**Results**

*Irradiation under grazing angle of incidence*

First we exploited the possibility to locally analyze our samples by means of SPM to characterize the divergence of the UNILAC beam in our irradiation chamber. This is an important parameter for angle dependent experiments. For this experiment we irradiated single crystals of strontium titanate ($SrTiO_3$) with $^{136}Xe^{21+}$ and 4.8 MeV/u under an angle of incidence of around 2°. $SrTiO_3$ is a good system for our purpose as it has been very well characterized with respect to grazing incidence irradiation damage [7,17]. At this ion energy range with a stopping power of around 29 keV/nm (SRIM calculation with an assumed density for $SrTiO_3$ of 5.11 g/cm$^3$), one ion impact causes one irradiation event on the surface [24]. Under grazing incidence each impact gives rise to the creation of a chain of nanosized hillocks, easily detectable with an AFM. The nominal fluence for the irradiation here was $1.99 \cdot 10^{10} \pm 10\%$ ions/cm$^2$. This value was measured at an aperture in the beamline and was calibrated by the second Faraday cup in the M1-beamline. With an assumed efficiency of one, this fluence should lead to around 7 ion induced features per µm$^2$ which should be aligned along the direction of the ion beam. Figure 6 shows a typical AFM image of the irradiated $SrTiO_3$ surface. Chains of hillocks with an average length of 690 nm ± 187 nm (average over 7 images, each frame size was 2.5 x 2.5 µm$^2$) corresponding to an average fluence of 5.1 ± 0.8 features/µm² can be observed. The difference in the number of events/µm² can be explained by a misalignment of the angle of incidence or with an error of the measured nominal fluence. By comparing the fluence $F_{cup}$, which was measured in the Faraday cup, and the events per element of area $F_{event}$ it is possible to estimate the real angle of incidence $\alpha$:

$$\alpha = \arcsin(\frac{F_{event}}{F_{cup}})$$

The true angle of incidence was thus $1.5° \pm 0.4°$.

For a quantitative estimation of the beam divergence the chains of hillocks can now be used as rulers, as each chain represents the track of an ion. As one can see from figure 6 the chains of hillocks are aligned along the beam direction and appear parallel to each other. However, a few tracks show some small misalignment with respect to nominal beam direction. For the analysis we overlaid all visible tracks with lines in the AFM images and measured the angles between these lines and a reference line. From this we can calculate the lateral beam divergence to be $0.6° \pm 0.2°$. One can estimate the maximum distance between two tracks with different lateral angles starting from the same impact point by simple trigonometry. The distance between two chains of hillocks with a length of 690 nm and a lateral misalignment of 0.6° will be around 7 nm. Thus, the beam divergence becomes very important for experiments with ultra-grazing angles of incidence. This can also be seen in the large error bars of the length of chains of hillocks determined in former experiments under grazing angle of incidence [7], where the vertical divergence of the beam obviously causes a significant variation in the track length. For chains of hillocks of some tenth of nm this divergence is however negligible.

Unexpectedly, we observed a novel ion-induced feature in this experiment, namely a very shallow ($0.3 \pm 0.1$ nm) and narrow ($9 \pm 2$ nm) rift in front (seen from the beam direction) of the chain of hillocks (see inset figure 6). These rifts occurred in 66% of the events and had a length of 186 nm $\pm$ 50 nm. In the remaining 34% features a rift cannot unambiguously identified. This can have two reasons. In some cases the ion induces surface tracks which are located close to each other, thus the length of the track is measurable easy enough but a 0.3 nm shallow rift is simply not distinguishable right next to a 6 nm high hillock. In addition, a rift may simply be covered up again by the second track. A second reason is the image size that was chosen for these first experiments. In a frame of 2.5 x 2.5 µm² with a typical number of 1000 pixels per line, one pixel has the size of 6.25 nm². That is, if a feature is smaller in width than 2.5 nm, it will not be imaged correctly. In addition, an AFM image is always a convolution of the surface structure and the tip (typically with a radius of 10 nm) which may further obscure small features.

The creation of rifts by SHI has just recently been demonstrated in the case of SiC [3] but for SrTiO$_3$ they have not been described before. The preliminary data collected so far prevents us from giving a detailed interpretation but the rifts demonstrate nicely the excellent imaging capacity of our SPM setup as well as the necessity of working in the cleanest conditions possible as these shallow features may be easily covered by adsorbates from ambient conditions making them undetectable by ambient AFM.

*MeV-SIMS/SNMS*

To demonstrate the capabilities of the TOF mass spectrometer, probably the first choice of test samples are ionic crystals, since these are well known to exhibit large electronic sputtering yields. An example of the raw data taken on a KBr sample is depicted in figure 7, which shows a screen shot taken from the data acquisition software using a total number of 36 UNILAC pulse cycles. In this experiment, the UNILAC was running with a pulse length of about 1 ms at a repetition rate of about 2 Hz, so that the acquisition of the entire data set took only about 18 s. The data acquisition frequency was set to 1 kHz, and the first acquisition in every gate was timed to occur about 120 µs after the start of the UNILAC trigger pulse. With these settings, only two sweeps fit in a gate window of 2 ms duration, rendering the number of acquired sweeps (in the software called "reps") in the MeV-SIMS spectrum equal to that in the MeV-SNMS spectrum. The keV gate window width was set to 20 ms, thereby accommodating one SNMS (with the laser fired) and five SIMS sweeps (without the laser fired) per UNILAC pulse cycle.

First, and most importantly, it is seen that the synchronization between hard- and software apparently works. Neither the residual gas spectrum nor the blank spectrum exhibit any signal apart from the baseline, which would not be the case if only one single sweep with either one of the ion beams switched on would have been erroneously sorted into these spectra. The acquired MeV-SNMS spectrum exhibits huge signals, which are large enough to necessitate a significant reduction of the MCP gain voltage to 1850 V in order to avoid detector saturation. The positive MeV-SIMS spectrum is clearly discernable, but shows much less intensity and a significantly lower number of peaks. With the same detector settings, both keV spectra exhibit essentially only one single peak at mass 40, which corresponds to the main isotope of potassium. Note that all displayed spectra are normalized such as to display the true relative intensity ratios, with the exception that the intensity scale of both MeV spectra has been reduced by a factor 25 with respect to the keV and blank spectra. It is evident that electronic sputtering of KBr produces a large number of atoms and clusters that are emitted in the neutral charge state. Figure 8 shows a detailed view of the measured spectra, revealing large contributions of K atoms and $KBr_n$ clusters in the sputtered flux. In order to unravel sputtered neutral from secondary ion signals, one has to keep in mind that the SNMS spectrum contains the SIMS spectrum as a background. This is illustrated in figure 9, where the SIMS spectrum is superimposed to the SNMS data as a white line. For reasons outlined below, the SIMS trace had to be multiplied by a factor two in order to fit to the measured SNMS spectrum. One immediate observation is that SIMS and SNMS signals exhibit different peak shapes, with the post-ionized neutral peaks being sharper than those of the respective secondary ions. This finding is due to the fact that the post-ionization laser was focused and did not entirely illuminate the sensitive volume of the mass spectrometer. Sputtered neutral particles are therefore only detected from the central part of the detection volume corresponding to good flight time focusing conditions, while secondary ions extracted from the outer parts of the sensitive volume contribute to the wings of the flight time peak. In any case, the peaks above the white SIMS background line unambiguously correspond to post-ionized sputtered neutral particles. It is seen that the large majority of the detected $K_2Br^+$ and practically all of the detected $KBr^+$ clusters are emitted in the neutral state, and even the measured $K^+$ signal exhibits a significant contribution from emitted neutral K atoms. Particularly the latter finding is interesting, since it is markedly different from keV sputtering, where practically all sputtered alkali atoms like Na or K are emitted as positive ions (see figure 7).

NaCl and $SrTiO_3$ as samples. In figure 10a cutout of a keV-SIMS and a MeV-SIMS spectrum of sodium chloride (NaCl) are shown. The upper panel shows the spectrum collected under 5 keV $Ar^+$ bombardment, the lower one shows the spectrum collected under 4.8 MeV/u $^{197}Au^{26+}$ bombardment. The spectra show the signals for the sodium monomer at mass 23 u and the dimer at mass 46 u. In the spectrum collected with the particle accelerator as primary ion source shows additional peaks compared to the keV spectrum. These additional peaks appear due to the special data acquisition mode developed for measurements with the long UNILAC pulses as described above. The spectrometer is tuned in a way that ions are time focused onto the detector which originates from a volume in a distance $d$ above the surface. The ions from that volume are accelerated by an electric field of the strength $|\vec{E}_{extraction}| = U_{extraction}/(d-h)$, where $h$ is the distance between the surface of the sample and the entrance of the spectrometer. $U_{extrcation}$ is the extraction voltage. When the extraction voltage is ramped up the time focus conditions are fulfilled for ions, which filled up the extraction volume. As described above in IIIb, the extraction voltage is ramped while the UNILAC is still bombarding the surface, leading to a constant flux of ions from the surface. The ions sputtered during the ongoing bombardment are accelerated away from the surface into the spectrometer in the full field strength of the extraction field, leading to an unfocussed DC background in the spectrum. This background appears in a lift in the baseline of the spectrum on the right hand side of each peak with the same width as the extraction pulse. Additional features in the spectra collected with the UNILAC are further peaks in the spectrum not corresponding to the sample material. The extra peaks at mass 32 u and 58 u can be explained by the extraction process. Within the ongoing bombardment of the surface by swift heavy ions the extracted ions are continuously accelerated while the extraction voltage is switched on. When the extraction voltage is ramped down at the end of the extraction pulse, the time focus conditions are fulfilled for particles from a volume $h - d$ above the surface. For these ions the accelerating field strength is the same as for ions inside the sensitive volume in the distance $d$ above the surface when the extraction voltage is ramped up. All theses extra peaks appear delayed to the mass peaks by the width of the extraction pulse. It is not possible to create a very short extraction pulse to suppress this effect, because the minimum width of the extraction pulse is determined by the time an ion of a distinct mass needs to travel the way from its starting point into the spectrometer. Assuming $h \ll d$ the minimum extraction time can be calculated as

$$t^{min}_{extraction} \approx \sqrt{\frac{2d^2}{U_{extraction}} \cdot \frac{m}{q}}$$

Is the width of the extraction pulse chosen too short, the higher mass ions will be suppressed in the spectrum. The mass peaks of the extra peaks is broadened thus means the mass resolution is reduced because the extraction potential is not ramped to zero immediately and the time the particles need to enter the spectrometer is reduced. Both reduce the mass resolution.

The identification of the extra peaks can be easily done by altering the width of the extraction pulse and observing the shift of the extra peaks. The peaks contributing to a real constituent of the sample do not shift with the width of the extraction pulse but the extra peaks shift proportional to the width of the extraction pulse.

*Time behavior of the UNILAC pulse*

Using the interleaved data acquisition protocol described above, the profile of the ion pulse generated by the particle accelerator is of great importance for the interpretation of the spectra. Especially for a quantitative comparison of the secondary ion and neutral yields in order to calculate the ionization probability of the sputtered particles, the protocol implicitly assumes identical experimental conditions during acquisition of SNMS and SIMS spectra. While all other conditions remain unchanged, there is, however, a fundamental difference between both spectra, since they are being acquired at different, although constant, times during an UNILAC pulse. Since measured mass spectral signals are directly proportional to the projectile beam current, this implicitly assumes that current to be the same at all times during a pulse, corresponding to an ideally rectangular pulse profile with a constant ion flux onto the surface throughout the pulse. The real pulse profile differs significantly from that ideal profile. As a consequence, it is not sufficient to assume a reproducible pulse profile that can be measured once and used to correct the relative intensities of spectra taken at different times during the pulse. In fact, we have found that the pulse profile may change on an hourly basis, making it necessary to find a way to determine its actual shape for every experiment separately. Due to the proportionality between measured signals and the projectile flux, this can be done in a direct way using the TOF-SNMS/SIMS data itself. A relatively easy way to achieve this would be to use the SNMS signal and vary the delay between UNILAC trigger and data acquisition. Depending on the desired time resolution, this method works, but it requires the sequential acquisition of a full series of SNMS spectra and therefore only reveals information averaged over many UNILAC pulses. In addition, its success relies on the assumption that all other experimental conditions remain exactly constant during the entire acquisition time, a prerequisite which is hard to ensure at a large scale accelerator facility. We have therefore implemented another method which makes use of SIMS data and allows measuring the pulse profile on a pulse-by-pulse basis. For that purpose, we increase the data acquisition rate of the transient digitizer by only collecting a short portion of the flight time spectrum, thereby mapping a single prominent secondary ion peak such as, for instance, the $K^+$ signal in the spectrum in figure 7. This way, a maximum acquisition rate of about 5 kHz can be achieved, which determines a time resolution of about 200µs for the measurement of the pulse profile. Similar to the interleaved data acquisition described above, the data collected at each point in time during one UNILAC pulse is stored in separate spectra, which can either be read out on a sweep-by-sweep basis or summed from pulse to pulse. In principle, this allows to measure the temporal structure of a single accelerator pulse, so that fluctuations within the pulse-to-pulse statistics can be investigated. To further increase the time resolution, the entire data acquisition scheme can be shifted stepwise by increments of, say, 100 µs, thereby filling the gaps between subsequent data points.

An example of such a measurement is shown in figure 11. In this case, the Na$^+$ signal was monitored during bombardment of a Fluorenylmethyloxycarbonyl (FMOC) surface[1]. It is evident that a single UNILAC pulse can successfully be characterized using this technique. Moreover, it is seen that a single pulse profile may significantly deviate from that averaged over many pulses. The dotted line depicts an oscilloscope trace taken in the main control room of the accelerator facility after completion of the experiment, which displays the pulse shape measured there and averaged over many pulses. During the acquisition of the data depicted in figure 7, this pulse mapping technique was not yet available to us, so that we can only speculate that the apparent intensity difference between the MeV-SIMS spectrum and the clearly identifiable SIMS background in the MeV-SNMS spectrum must originate from a temporal variation of the primary ion current as well. Fitting the SIMS spectrum to the SIMS background in the SNMS spectrum as indicated in figure 9, we had to increase the intensity by a factor two. Note that the SNMS spectrum was always taken at the beginning, while the corresponding SIMS spectrum was always taken near the end of each UNILAC pulse. The data presented in figure 11 show that a variation of the projectile ion flux by about a factor two within the pulse is clearly possible, lending credit to our interpretation. It should be noted that the decrease of the ion current during the pulse as shown in figure 11 is not a general feature of the UNILAC, since we have also observed the opposite behavior at different times.

*Spectrum for the analysis of rifts*

In the last sections we demonstrated the serviceability of the single parts of this setup. The exceptional advantage of SHIPS is the synergy of an analysis of emitted material during the irradiation and a characterization of the surface structure after irradiation. For the first time we are able to connect sputtering of the surface with morphological changes.

Surface tracks may manifest themselves in many different forms. Material may protrude from the surface, or rifts are formed, or sometimes even no permanent changes are detected although transient changes may have taken place, as has been demonstrated for NaCl e.g. [25]. In all cases, the analysis of ejected particles during irradiation can help to understand how the surface track is actually formed. In the case of rifts the TOF will give direct information about the missing material, in case of a seemingly unchanged surface transient processes may be revealed, and in the case of protruding material a detailed analysis

---

[1] This data acquisition method was not yet available when the data of figure 7 were taken

of particle distributions may be used to identify hidden processes such as Coulomb explosion. As discussed in section "*Irradiation under grazing angle of incidence*", a $SrTiO_3$ surface shows the removal of material in front of the chains of hillocks in form of a rift after irradiation under gracing angle of incidence. In contrast titanium dioxide does not show such a rift formation. We therefore took advantage of our new setup to see if the analysis of the sputtered particles with respect to their mass and composition gives evidence for the rift formation. Figure 12 shows a MeV-SIMS of $SrTiO_3$ surface under 4.8 MeV/u $^{197}Au^{26+}$ bombardment. It can be clearly seen that under SHI irradiation indeed significant particle emission takes place. The main peaks in the spectrum can be attributed to Sr and its oxides, while Ti and TiO ions are detected only to a very small amount. This difference might provide insight into the sputtering and rift formation mechanism but further data like SNMS spectra to detect possible neutral particles will be necessary in order to do so in future experiments.

**Conclusion**

We have installed a new UHV setup for surface science and material research at the M-branch of GSI called SHIPS. With the new setup it is now possible to prepare and irradiate samples under well-defined conditions and analyze them in-situ by complementary surface sensitive methods. We can investigate permanent structural modifications after irradiation by means of SPM and at the same time study the particle ejection process in-situ during irradiation via TOF-SNMS/SIMS, which allows analyzing the composition of the plume of sputtered material up to masses above 1000 u. The particular strength of the instrument developed here is the fact that not only sputtered ions, but also the neutral components of the sputtered flux can be investigated, a feature that is worldwide unique though extremely important since the neutral components form the vast majority of the ejected material. The experiment therefore gives the opportunity to investigate the sputtering process in the electronic stopping regime induced by the impact of swift heavy ions with kinetic energies around 1 GeV in direct comparison to the linear cascade sputtering process occurring in the nuclear stopping regime induced. The scanning probe microscope allows a detailed in-situ analysis of the ion-induced permanent modifications with the highest possible resolution. The unique possibility to combine data on ejected particles with the local probing of ion-induced material changes will provide a big help to understand the physical mechanisms at the origin of SHI induced material modifications.

**Acknowledgments**

We want to express our thanks to the Federal Ministry of Education and Research for the financial support, to Dr. T. Peters, Dr. O. Picht, and F. Weiss for their help and work during the starting phase of the project and to A. Sigmund and W. Saure for their good technical support during the whole project.

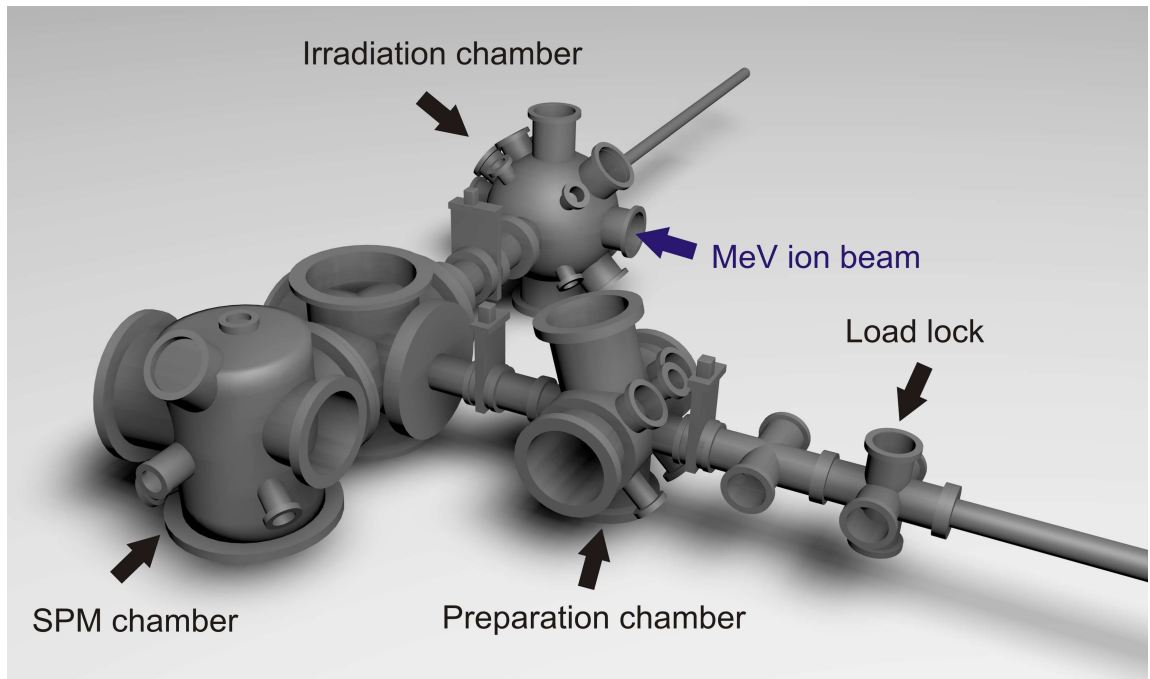

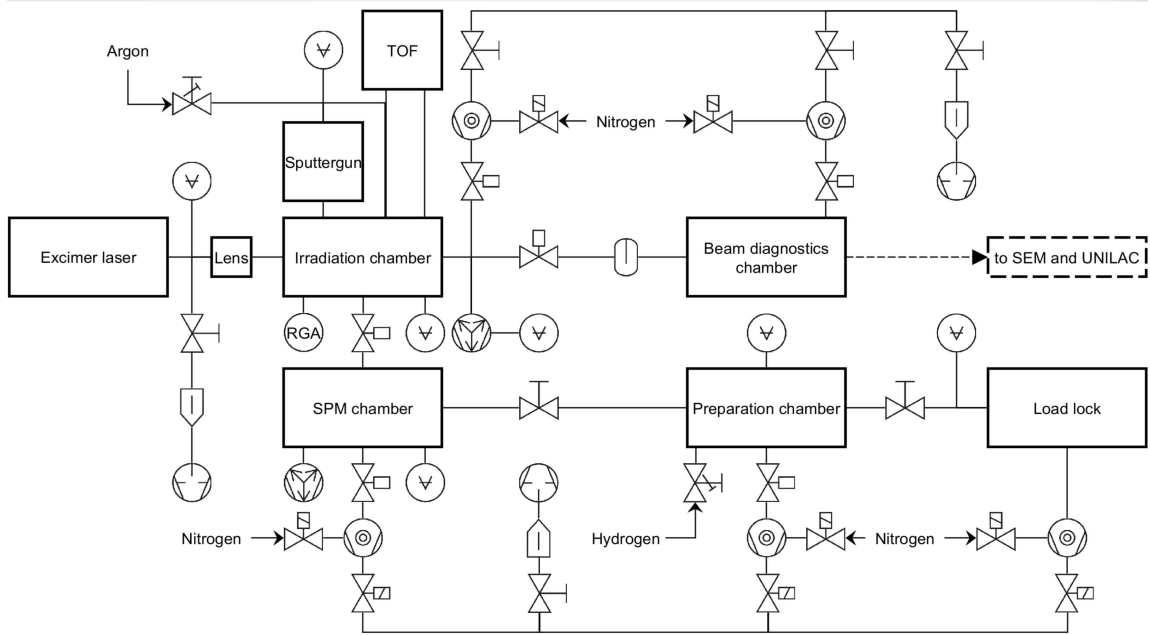

**Figure 1: Scheme of the vacuum system at the M1 beamline at GSI in Darmstadt, Germany. A 3D-graphic account with the different main chambers of the setup is shown in the upper panel. The lower panel shows a scheme of the different vacuum components in this setup.**

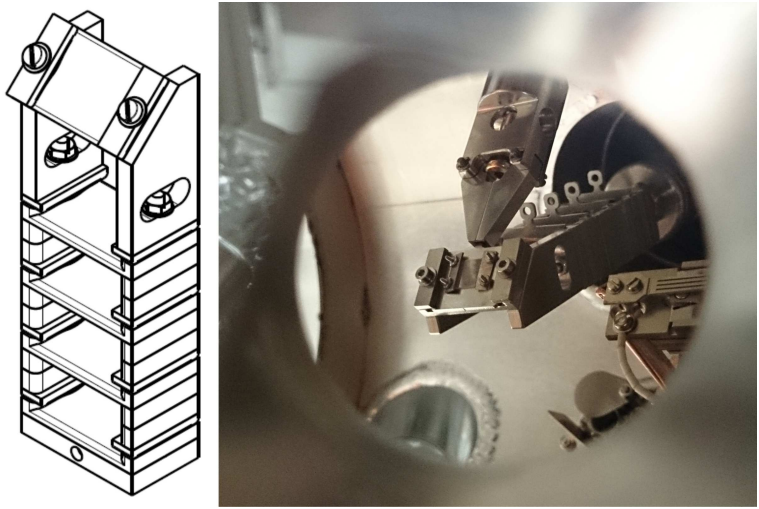

**Figure 2: Scheme of the transfer adapter and picture during the transfer process of the sample in the preparation chamber with the wobble stick**

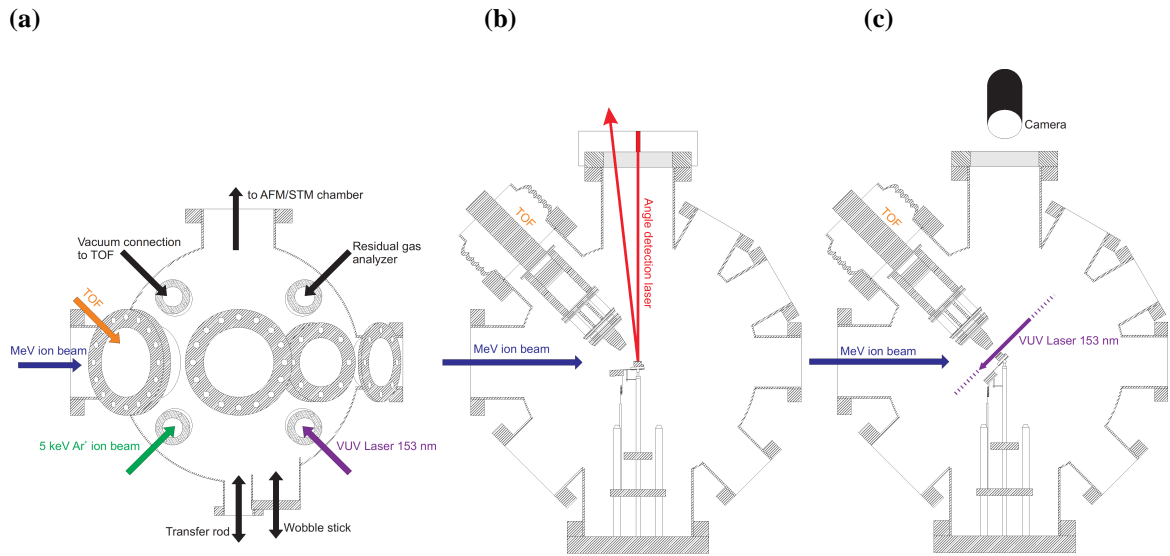

**Figure 3: (a) Top view of the irradiation chamber, with arrows to mark the configuration of the flanges (b) Side view of the irradiation chamber for experiments under grazing angle of incidence with a red laser diode to detect the tilting of the sample against the MeV ion beam and (c) for TOF-MS experiments with the sample under 45° towards the MeV ion beam.**

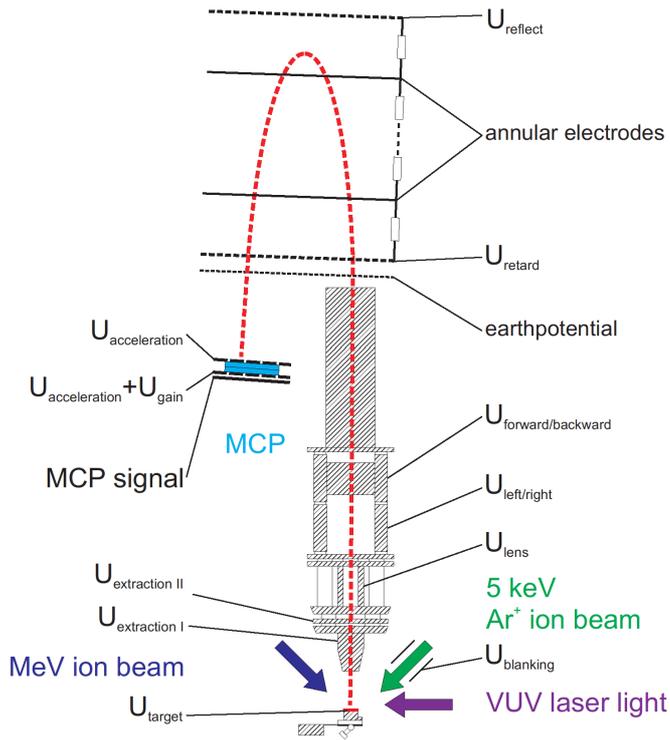

**Figure 4:** Schematic drawing of the TOF-MS, with marks to places where the different voltages are applied. The red dashed line stands for the trajectory of the accelerated ions from above the surface.

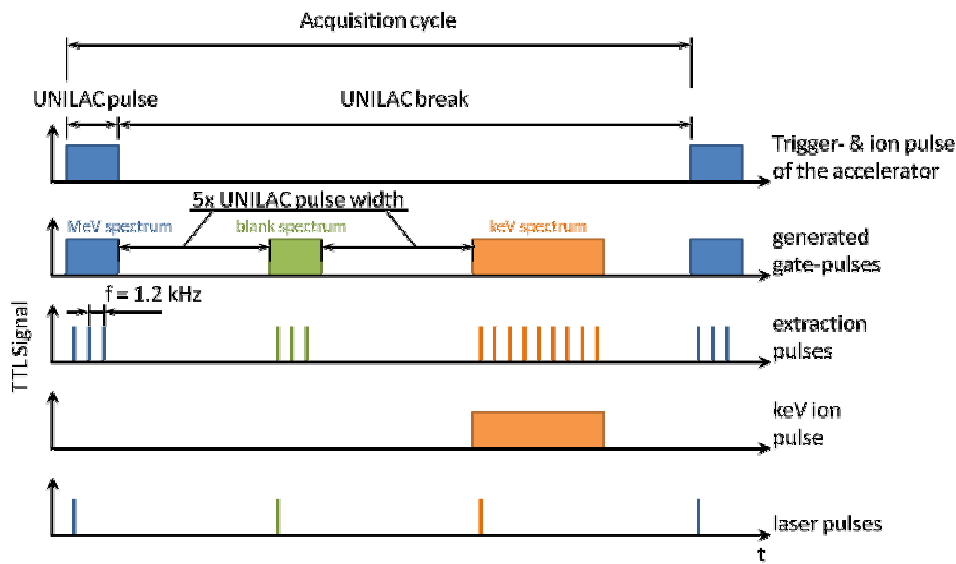

**Figure 5: Timing scheme of the MeV-SNMS/SIMS experiment. Within one acquisition cycle six spectra are acquired: two spectra with MeV ion bombardment (MeV), two without any ion bombardment (Blank) and two with 5 keV Ar+ bombardment (keV). For each gate in the first extraction cycle the laser is triggered (SNMS) for all other extraction within the same gate the laser trigger is suppressed (SIMS). (a) UNILAC trigger pulse provided by the GSI main control room (b) Artificially generated gate pulses for gating the delay generator (c) extraction pulses generated by the delay generator (d) Trigger signal for firing the laser, suppressed for SIMS measurements (e) Trigger signal for firing the 5 keV Ar+ ion gun**

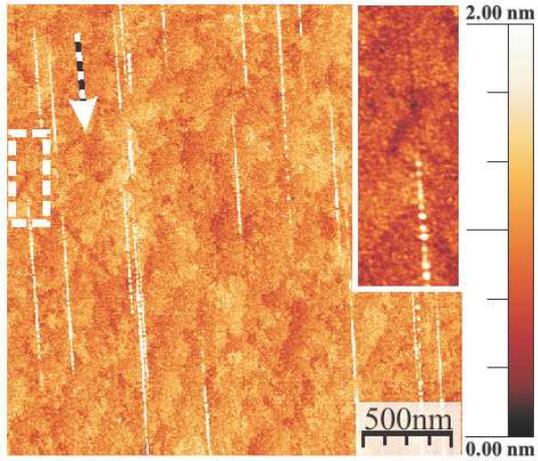

**Figure 6:** AFM-Image of a SrTiO$_3$ surface ($df$ = -22 Hz). The sample was irradiated under an angle of θ = 2° ± 0.6° with $^{136}$Xe$^{21+}$ and an energy of 653 MeV. The stopping power was 29 keV/nm. The arrow marks the direction of the ion beam. The white dashed frame designates the area of the inlet. The bright lines in the image are chains of hillock. At the beginning of the chain of hillocks narrow (9 nm) and shallow (0.3 nm) rift can be observed, that is clearly seen in the inlet.

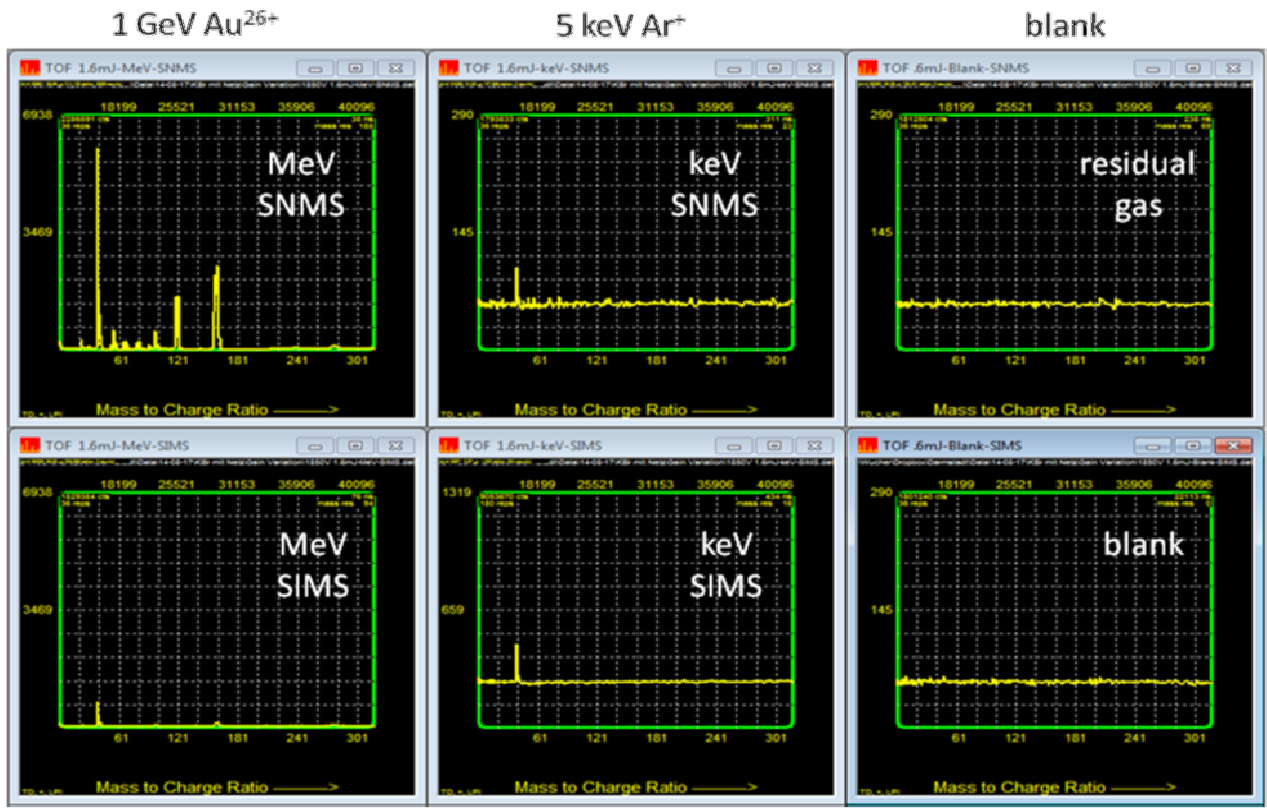

**Figure 7: Screen shot of TOF-SNMS/SIMS measurement taken on a KBr crystal covered with a grid (mesh size : 228 µm, wire diameter 25 µm) under bombardment with 4.8 MeV/u $Au^{26+}$ ions ("MeV-SNMS/SIMS") or 5 keV $Ar^+$ ions ("keV_SNMS/SIMS"), respectively, using the newly developed interleaved data acquisition protocol. Postionization laser: 157 nm, 8 ns, 1.6 mJ/pulse. The spectrum labeled "residual gas" was taken with the laser alone, while the spectrum labeled "blank" is the control without ion bombardment and laser.**

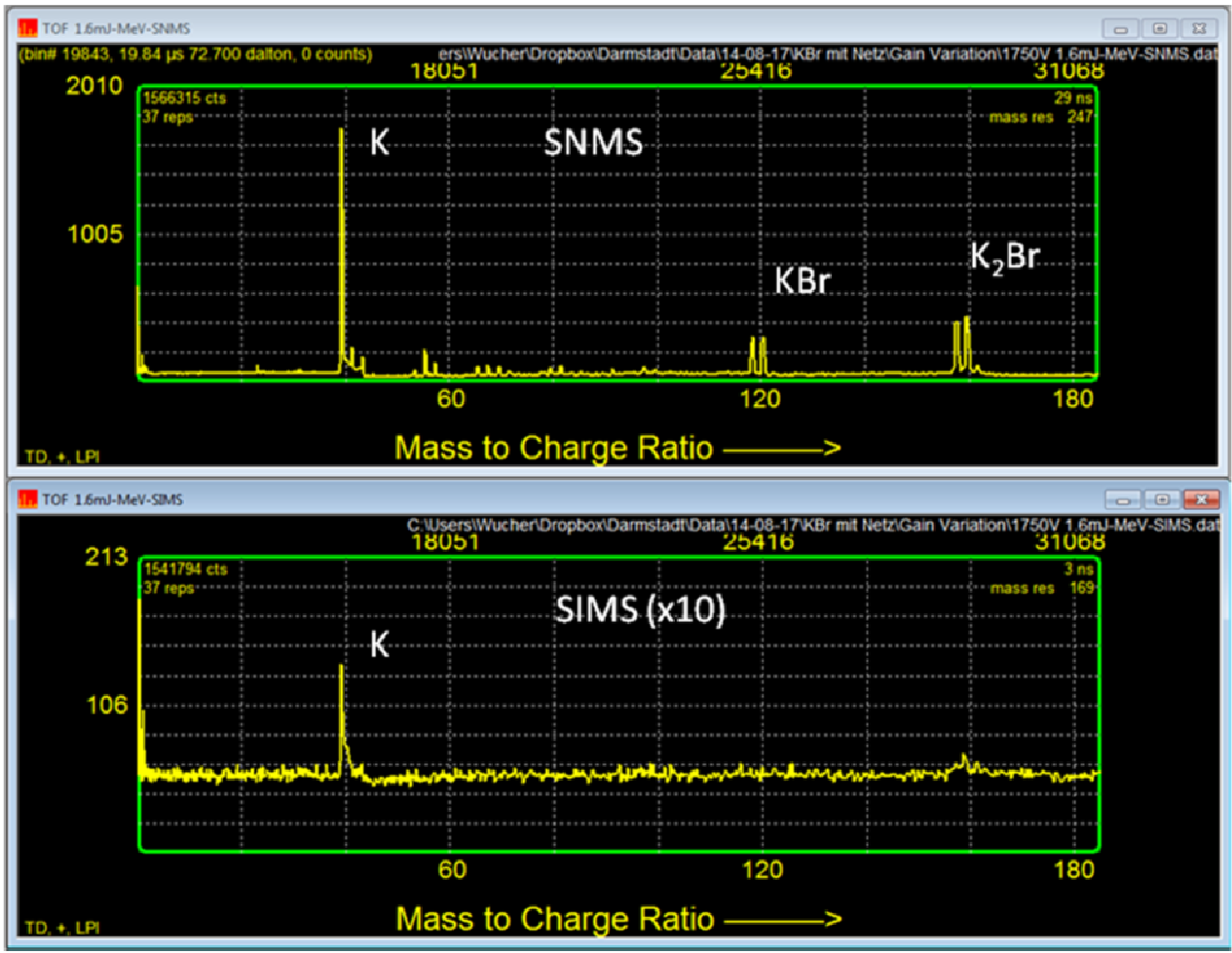

**Figure 8: Same data as displayed in figure 7 but now restricted to the SNMS and SIMS spectra measured under 4.8 MeV/u Au$^{26+}$ ion bombardment.**

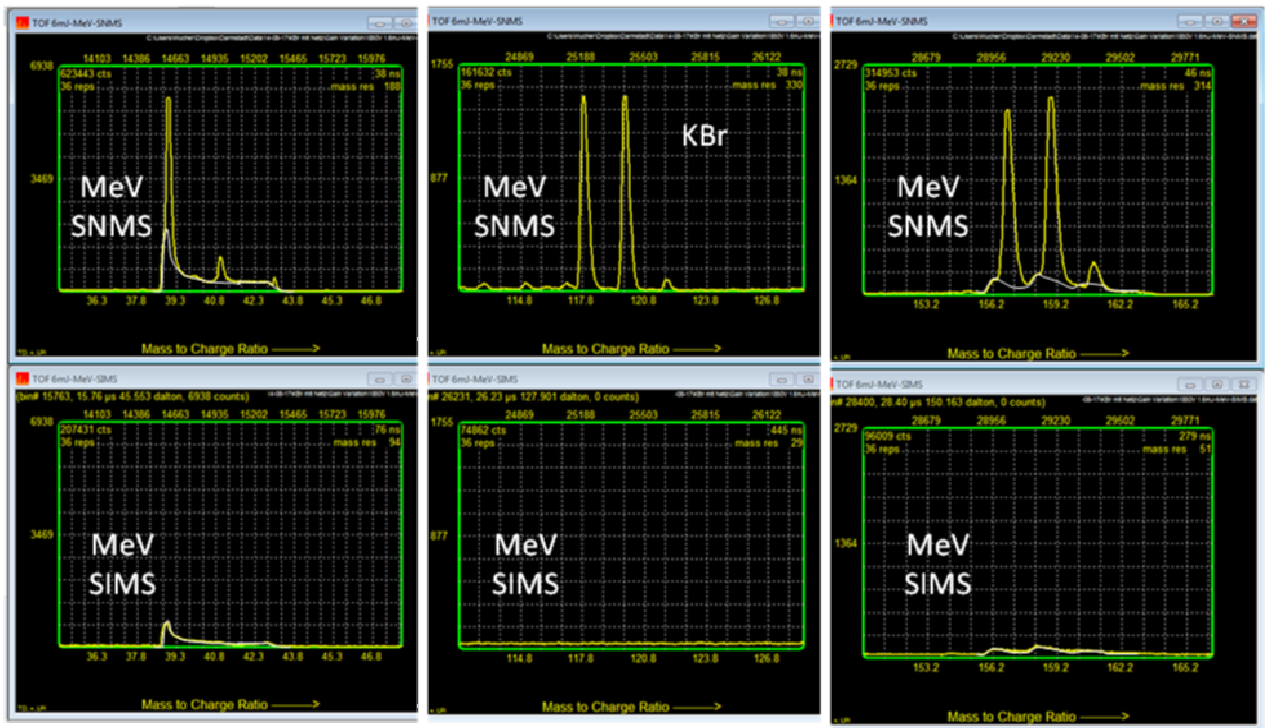

**Figure 9:** Same data as displayed in figure 7, but now restricted to the SNMS and SIMS spectra measured under 4.8 MeV/u Au$^{26+}$ ion bombardment.

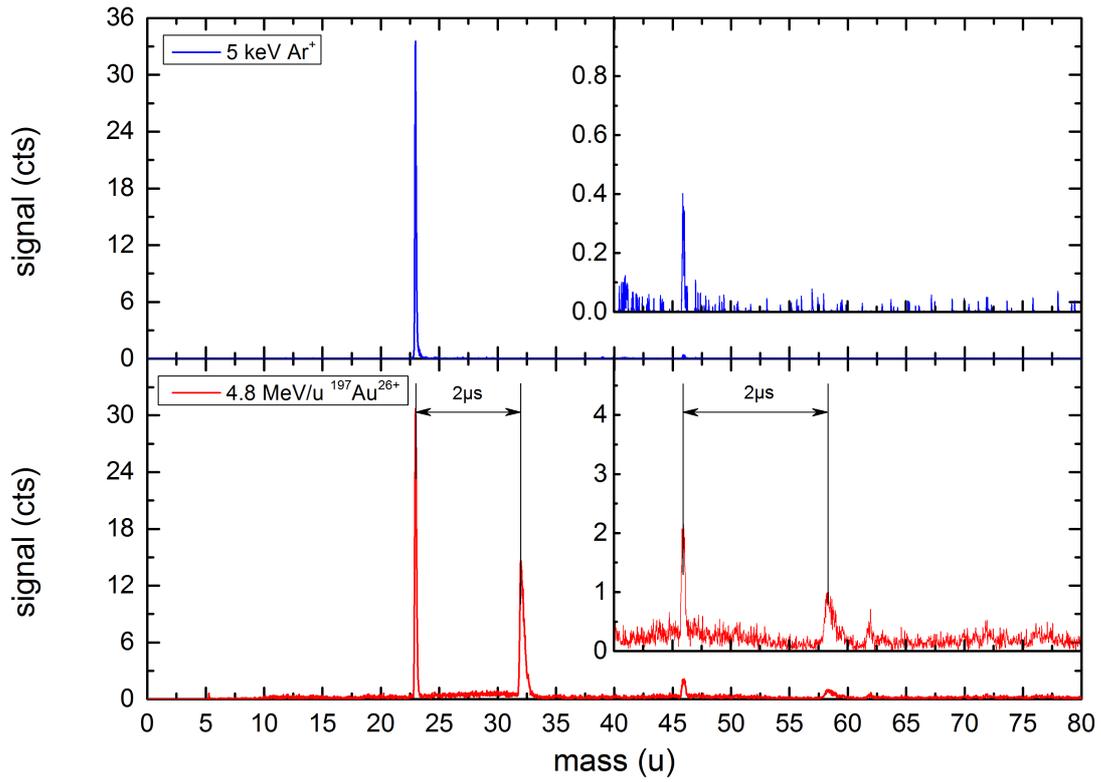

Figure 10: (a) Example for a common "delayed extraction" TOF-MS measurement of NaCl crystal with keV ions and (b) for an irradiation with swift heavy ion measured with the new developed "interleaved extraction"-method. In both cases the used length of the extraction pulse width is 2 μs.

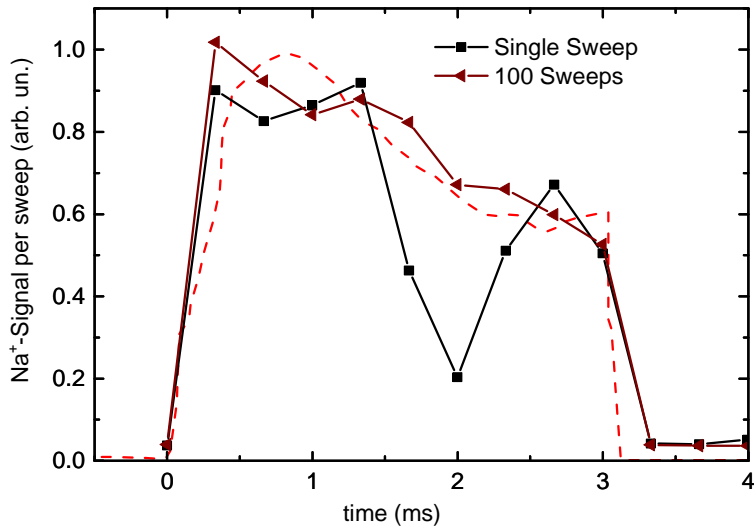

**Figure 11: Temporal profile of the UNILAC pulse measured by the pulse mapping technique described in the text. The dots represent the measured Na+ secondary ion signal as a function of time after the start of the UNILAC trigger pulse. The dashed line represents an oscilloscope trace retrieved from the main control room after completion of the experiment.**

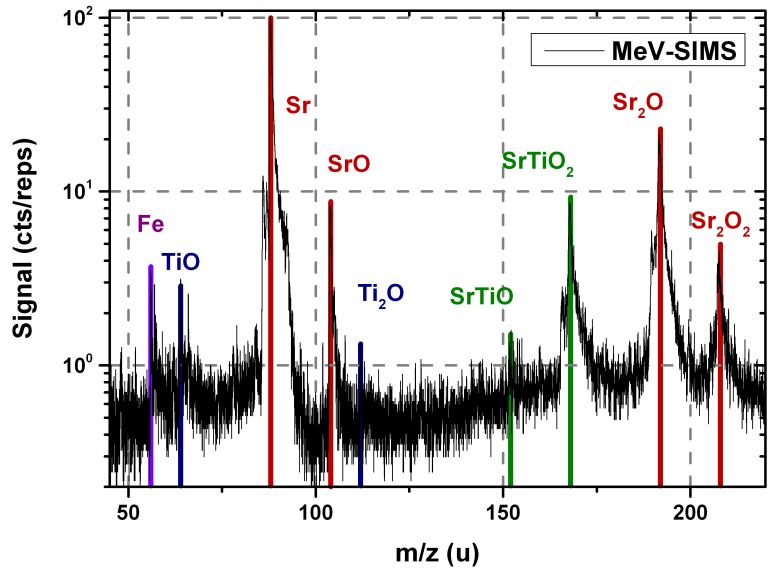

Figure 11: Secondary-ion-mass-spectrum of a $SrTiO_3$ sample irradiated with 4.8 MeV/u $^{197}Au^{26+}$. The single peaks are collated to different elements and composites.

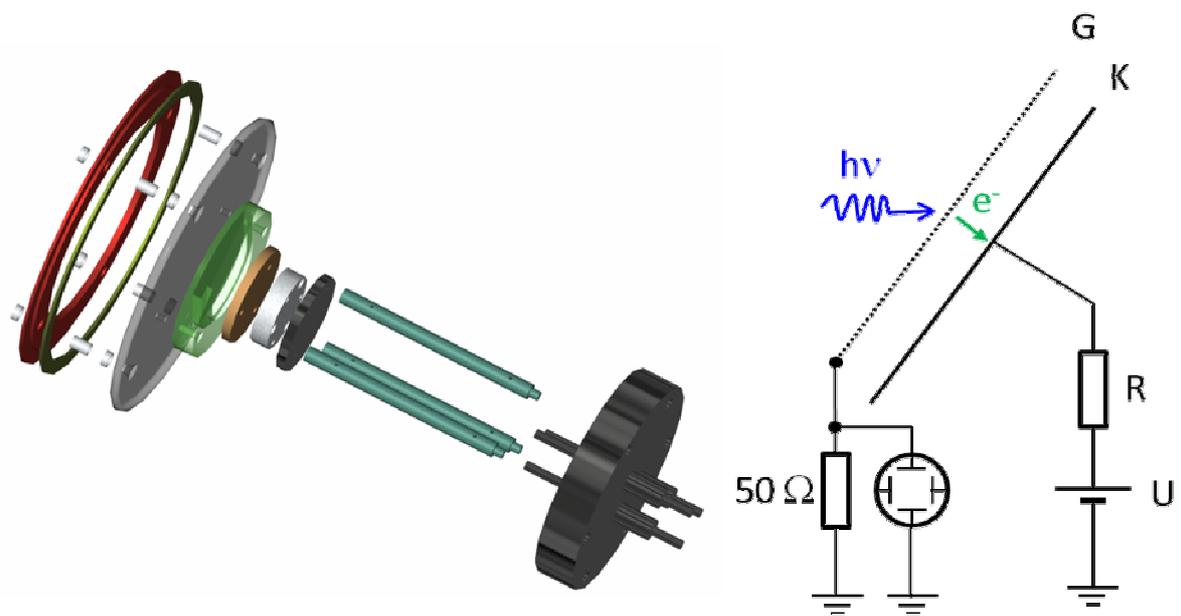

**Figure S 1: Mechanical design and schematic working principle of photoelectric detector to monitor the VUV laser pulse. [22]**

# References

[1] D. A. Young, Nature 182, 375-377; doi:10.1038/182375a0

[2] Quoc Hung Nguyen, Mubarak Ali, Reinhard Neumann, Wolfgang Ensinger; Sensors and Actuators B: Chemical, Volume 162, Issue 1, 20 February 2012, Pages 216-222; doi:10.1016/j.snb.2011.12.070

[3] O. Ochedowski, O. Osmani, M. Schade, B. Kleine Bussmann, B. Ban-d'Etat, H. Lebius, M. Schleberger, Nature Communications 5, 3913; doi:10.1038/ncomms4913

[4] O. Ochedowski, H. Bukowska, V. M. Freire Soler, L. Brökers, B. Ban-d'Etat, H. Lebius, M. Schleberger, Nucl. Instr. Meth., Volume 340, 2014, Pages 39–43; doi:10.1016/j.nimb.2014.07.037

[5] M. C. Ridgway, T. Bierschenk, R. Giulian, B. Afra, M. D. Rodriguez, L. L. Araujo, A. P. Byrne, N. Kirby, O. H. Pakarinen, F. Djurabekova, K. Nordlund, M. Schleberger, O. Osmani, N. Medvedev, B. Rethfeld, P. Kluth; Phys. Rev. Lett. 110, 245502; doi:10.1103/PhysRevLett.110.245502

[6] F. Aumayr, S. Facsko, A. El-Said, C. Trautmann, M. Schleberger; J. Phys.: Condens. Matter 23, 393001 (2011); doi:10.1088/0953-8984/23/39/393001

[7] E. Akcöltekin, S. Akcöltekin, O. Osmani, A. Duvenbeck, H. Lebius and M. Schleberger; New J. Phys. 10, 053007 (2008), doi:10.1088/1367-2630/10/5/053007

[8] M. Toulemonde, C. Dufour, E. Paumier, (1992) Phys. Rev. B 46, 14362, doi:10.1103/PhysRevB.46.14362

[9] S. Klaumünzer in Ion beam science by P. Sigmund, Matematisk-Fysiske Meddelelser, 52 (2006), pp. 293–328, ISBN 87-7304-330-3

[10] M. Toulemonde, C. Dufour, E. Paumier, (2006) Acta Physica Polonica A (109), S. 311; http://przyrbwn.icm.edu.pl/APP/PDF/109/a109z309.pdf

[11] G. Betz, K. Wien, International Journal of Mass Spectrometry and Ion Processes 140, 1, (1994) 1–110; doi:10.1016/0168-1176(94)04052-4

[12] H. Hijazi, H. Rothard, P. Boduch, I. Alzaher, A. Cassimi, F. Ropars, T. Been, J.M. Ramillon, H. Lebius, B. Ban-d'Etat, L.S. Farenzena, E.F. da Silveira, Eur. Phys. J. D (2012) 66: 68, doi:10.1140/epjd/e2012-20545-3

[13] H. Hijazi, H. Rothard, P. Boduch, I. Alzaher, Th. Langlinay, A. Cassimi, F. Ropars, T. Been, J.M. Ramillon, H. Lebius, B. Ban-d'Etat, L.S. Farenzena, E.F. da Silveira, Eur. Phys. J. D (2012) 66: 305, doi: 10.1140/epjd/e2012-30252-8

[14] R. E. Johnson, W.L. Brown, Nucl. Instr. Meth., 209/210 (1983) 469-476; doi:10.1016/0167-5087(83)90840-2

[15] P. K. Haff, Appl. Phys. Lett. 29, 473 (1976); doi:10.1063/1.89126

[16] R. L. Fleischer, P. B. Price, R. M. Walker, J. Appl. Phys. 36, 3645 (1965); doi:10.1063/1.1703059

[17] M. Karlusić, S. Akcöltekin, O. Osmani, I. Monnet, H. Lebius, M. Jaksić and Marika Schleberger, 2010 New J. Phys. 12 043009, doi:10.1088/1367-2630/12/4/043009

[18] C. Müller, M. Cranney, A. El-Said, N. Ishikawa, A. Iwase, M. Lang, R. Neumann, Nucl. Instr. Meth., Volume 191, Issues 1–4, 2002, Pages 246–250; doi:10.1016/S0168-583X(02)00569-4

[19] M Karlušić, R Kozubek, H Lebius, B Ban-d'Etat, R A Wilhelm, M Buljan, Z Siketić, F Scholz, T Meisch, M Jakšić, S Bernstorff, M Schleberger and B Šantić, J. Phys. D: Appl. Phys. 48(2015) 325304; doi: 10.1088/0022-3727/48/32/325304

[20] S. Amirthapandian, F. Schuchart, W. Bolse, Rev. Sci. Instrum. 81, 033702 (2010); doi: 10.1063/1.3316803

[21] A. Labuda, Y. Miyahara, L. Cockins, P. H. Grütter; Phys. Rev. B 84, 125433 (2011); doi: 10.1103/PhysRevB.84.125433

[22] L. Breuer, PhD thesis, 2015, Universität Duisburg-Essen

[23] M. Wahl, PhD thesis, 1995, Universität Kaiserslautern

[24] S. Akcöltekin, E. Akcöltekin, T. Roll, H. Lebius, M. Schleberger, Nucl. Instr. Meth., Volume 267, Issues 8–9, 2009, Pages 1386–1389; doi:10.1016/j.nimb.2009.01.156

[25] Ochedowski, S. Akcöltekin, B. Ban d´Etat, H. Lebius, and M. Schleberger; Nucl. Instr. Meth. B 314, 18 (2013); doi: 10.1016/j.nimb.2013.03.063